# Large-scale validation of an automatic EEG arousal detection algorithm using different heterogeneous databases

Diego Alvarez-Estevez[1], Isaac Fernández-Varela[2]

[1] Sleep center and clinical neurophysiology department, Haaglanden Medisch Centrum, The Hague, The Netherlands
[2] Computer Science department, University of A Coruña, A Coruña, Spain

\*\*\*

**Abstract**

Objective: To assess the validity of an automatic EEG arousal detection algorithm using large patient samples and different heterogeneous databases

Methods: Automatic scorings were confronted with results from human expert scorers on a total of 2768 full-night PSG recordings obtained from two different databases. Of them, 472 recordings were obtained during clinical routine at our sleep center, and were subdivided into two subgroups of 220 (HMC-S) and 252 (HMC-M) recordings each, attending to the procedure followed by the clinical expert during the visual review (semi-automatic or purely manual, respectively). In addition, 2296 recordings from the public SHHS-2 database were evaluated against the respective manual expert scorings.

Results: Event-by-event epoch-based validation resulted in an overall Cohen's kappa agreement κ = 0.600 (HMC-S), 0.559 (HMC-M), and 0.573 (SHHS2). Estimated inter-scorer variability on the datasets was, respectively, κ = 0.594, 0.561 and 0.543. Analyses of the corresponding Arousal Index scores showed associated automatic-human repeatability indices ranging in 0.693-0.771 (HMC-S), 0.646-0.791 (HMC-M), and 0.759-0.791 (SHHS2).

Conclusions: Large-scale validation of our automatic EEG arousal detector on different databases has shown robust performance and good generalization results comparable to the expected levels of human agreement. Special emphasis has been put on allowing reproducibility of the results and implementation of our method has been made accessible online as open source code.

## 1. Introduction

Electroencephalographic (EEG) arousals are transient events of the sleep EEG indicative of ongoing awakening activity. Specifically, according to the current clinical reference standards [1] EEG arousals are defined as abrupt shifts on the EEG frequency including alpha, theta and/or frequencies greater than 16 Hz (but not spindles) that last at least 3 seconds, with at least 10 seconds of stable sleep preceding the change. The scoring of arousals during the Rapid Eye Movement (REM) phase requires a concurrent increase in submental electromyogram (EMG) lasting at least one second.

Evidence supports EEG arousals as an important component of the sleep process, and their scoring during routine polysomnographic (PSG) examination is advised essential as a means of evaluating a subject's sleep continuity, and in order to give treatment and treatment response guidelines to practitioners [2].

Manual visual examination of the entire PSG for the scoring of these events is costly, due to the complexity and the amount of data involved. Given this context several works have explored the possibility of developing automatic analysis software to help the clinician during the scoring process [3] [4] [5] [6] [7] [8] [9] [10] [11] [12] [13] [14] [15] [16].

While some of the previous approaches have shown promising performance, validation methods are usually limited to relatively small (ranging 2-31 recordings), controlled, and mostly local and private datasets. It remains a question whether the detection capabilities of these algorithms generalize to larger samples, different databases, and perform well on a clinical (non-controlled) environment. Actually, the reality is that the grade of acceptability of these algorithms among the clinical community still remains low, being rarely used in the clinical practice.

In a recent work [16], we have presented a preliminary version of a method for automatic EEG arousal detection, obtaining good validation results on a controlled dataset of 22 PSG in-hospital recordings. Building upon this initial version, an updated algorithm has been developed and set it up to work in the clinical environment, where clinicians can choose whether to use it, or not, as a supportive scoring tool while reviewing their night recordings.

In this work we are presenting a large-scale validation on the performance of this updated approach. Validation has been carried out on a large sample of patients using our own sleep center database, but it has also extended by evaluating the algorithm over a large, external, and public database, namely the Sleep Heart Health Study (SHHS) [17]. On each case, the expected level of inter-scorer human variability has been estimated in order to contextualize the results of the analyses. Moreover, the source code of the algorithm has been published freely available on the internet for the research and the clinical community. To our knowledge this is the largest and more complete validation ever done of an algorithm of this kind.

## 2. Methods

### 2.1. Databases

The validation of the EEG arousal detection algorithm is performed using as reference data from two different and large databases. A first database is composed of clinical sleep recordings from our own sleep center (Haaglanden Medisch Centrum - HMC, The Hague, The Netherlands). Second, an external and public accessible source is used, namely the Sleep Heart Health Study (SHHS) database [17]. Each of the databases and the different derived datasets are described in detail in the following lines. A summary of resulting demographics and main PSG data are shown in Table 1.

#### 2.1.1. HMC

This data collection is composed of PSG recordings gathered retrospectively from the Haaglanden Medisch Centrum (The Hague, The Netherlands) sleep center database. PSG recordings were acquired in the course of common clinical practice, and thus did not subject people to any other treatment nor prescribed any additional behavior outside of the usual clinical procedures. Data were anonymized avoiding any possibility of individual patient identification. The study was carried out in full compliance with the corresponding applicable law and under the supervision of the local Medical Ethics Committee.

Patient signals were acquired using SOMNOscreen$^{TM}$ plus devices (SOMNOmedics, Germany) and digitized using the EDF+ format [18]. PSGs were afterwards analyzed offline by clinical experts in the course of common clinical practice. Manual scoring of the events included, among the common standard parameters, the annotation of sleep stages and of EEG arousals. All the procedures were performed according to the standard AASM guidelines [1]. As a homogenization criterion we required the recordings to contain at least 4 hours of Total Sleep Time (TST) after clinical scoring. To match a scenario as close as possible to the real working conditions, no further attempt was made to filter out, or to reject any recording, due to specific patient conditions or to poor signal quality. The only required condition was that the recording had been accepted by the clinicians for the manual scoring of EEG arousals.

In total, the sample included 472 recordings from patients visiting our center between April and October 2017. Data included both 24-h ambulatory (APSG, n=352) and in-hospital night (HPSG, n=143) recordings.

From this database two separated datasets were arranged as described below.

HMC-S:
A dataset containing 220 clinical recordings out of the original 472. Specifically, the dataset is composed of 176 APSGs and 45 HPSGs. PSG recordings from this dataset were clinically scored using a semi-automatic approach. First, the clinician used an automatic scoring algorithm for the detection of EEG arousals, and then, on a second pass, the scorer reviewed de results of the automatic scoring replacing, adding or deleting events where necessary. The automatic scoring algorithm used for this purpose was a previous version of the current approach which was described in detail in [16].

Table 1. Summary of demographic data and main PSG characteristics for the different datasets. Data are indicated as mean ± standard deviation; n = number of recordings, M = Males, F = Females, ArI = Arousal Index, AHI = Apnea-Hypopnea Index

| Dataset | n | Age | Gender | ArI | AHI |
|---|---|---|---|---|---|
| HMC-S | 220 | 52.99±14.33 | 137 M (62%) / 83 F (38%) | 12.85±7.86 | 13.33±15.28 |
| HMC-M | 252 | 51.58±16.22 | 133 M (53%) / 119 F (47%) | 12.45±10.48 | 14.83±21.03 |
| SHHS2 | 2296 | 67.41±10.03 | 1026 M (45%) / 1270 F (55%) | 12.91±7.02 | 16.25±15.64 |

HMC-M:

This dataset is composed of 252 recordings out of the original 472, split into 176 APSGs and 98 HPSGs. Recordings from this dataset were scored following the classical clinical routine, i.e. purely-manual without any support of automatic scoring.

2.1.2. SHHS

The Sleep Heart Health Study (SHHS) is a multi-center cohort study implemented by the National Heart Lung & Blood Institute to determine the cardiovascular and other consequences of sleep-disordered breathing. This database is available online upon permission at the National Sleep Research Resource (NSRR) [19] [20]. More information about the rationale, design, and protocol of the SHHS study can be found in the dedicated NSRR section [21] and in the literature [17] [22]. A sample of participants who met the SHHS inclusion criteria (age 40 years or older; no history of treatment of sleep apnea; no tracheostomy; no current home oxygen therapy) was invited to participate in the baseline examination of the SHHS, which included an initial polysomnogram (SHHS-1). In all, 6441 individuals were enrolled between November 1, 1995, and January 31, 1998. During exam cycle 3 (January 2001- June 2003), a second polysomnogram (SHHS-2) was obtained in 3295 of the participants. Raw PSG data are available at NSSR for 5793 subjects within SHHS-1, and for 2651 subjects within SHHS-2.

Polysomnograms were obtained in an unattended setting, usually at home, by trained and certified technicians. Specifications for the full montage settings can be found in the corresponding section at the NSRR website [21]. Scoring of sleep stages in SHHS is based on the R&K guidelines [23]. Note that in SHHS, however, no attempt was made to distinguish Stage 3 from Stage 4 which are combined into a single "Deep Sleep" category, similarly as in the current AASM standards [24] [1]. Scoring of arousals was done following the ASDA1992 manual [25]. Full specification of all the scoring criteria, as well as quality control procedures for the SHHS study, can be found in the Reading Center Manual of Operations [26].

From SHHS, the SHHS-2 dataset was used as the reference to validate our EEG arousal detection algorithm. Out of the 2561 PSGs available at NSRR, a total of 2492 recordings were selected after excluding those that did not match the general SHHS2 v3 signal montage [27], or for which no attempt to score EEG arousals was performed by the SHHS scorers. From this subset, 60 recordings were further excluded due to complete technical failure (complete absence of workable EEG and/or EMG signal during the whole recording) leading to 2433 recordings left. No further attempt was made to filter out, or reject any recording, due to poor signal quality conditions. Similarly as for HMC datasets, the selection excluded recordings for which TST < 4 hours. In total 2296 recordings were finally included for the validation of our algorithm.

2.2. Algorithm overview

The current version of the EEG arousals detection algorithm is largely based on a previous version which was described and validated elsewhere [16]. The current updated version is an evolution, although it preserves the same original philosophy of simplicity and robustness.

The algorithm works using just one EEG and one EMG chin derivations. The use of an additional ECG channel for the removal of ECG artifacts is optional. The method for ECG artifact removal is based on adaptive filtering and it has been described in detail elsewhere [28] [16].

It is a multistage method that consists of a first signal preprocessing step (digital Notch filtering in both signals, with optional adaptive ECG artifact removal, and high-pass filtering in the EMG), followed by detection of candidate events based on frequency changes in the EEG (power content analysis in the alpha (8–12 Hz) and in the beta (>16 Hz) bands). As a result of these steps candidate arousal regions are identified signaling the presence of EEG arousal activity. The analysis proceeds by pattern matching this activity using individual EEG and EMG relevant features. Eventually, candidate EEG arousals can be merged together, or the initial arousal region can be adjusted, usually by extending the corresponding event's offset, if recognizable arousal activity follows. The detection of EEG arousal events involves different subroutines including (i) EEG power-based, (ii) EEG amplitude-based, and (iii) EMG amplitude-based pattern recognition. Finally, false positives are discarded after examining each candidate event within the context of the accepted clinical definitions, including (i) its adequacy to the standard event duration constraints, (ii) the absence of sleep spindle activity, (iii) the absence of EMG activity during REM periods, and (iv) the presence of (at least) 10s of stable sleep preceding the onset of the event.

Adaptations to the algorithm were necessary on the first term to support the possibility of different montage configurations, signals sampling rates, and filtering settings (e.g. mains interference occurs in America at 60 Hz while at 50 Hz in Europe). In addition, detection thresholds were modified to increase the sensitivity for detection of alpha-prevalent arousal events, and to achieve better discrimination of sleep spindle activity. A number of processing steps were also simplified, namely regarding the EEG power-based pattern recognition (skipping the first false detection check, and discarding the whole procedure in the case of alpha arousals) and the detection of concurrent EMG activity. Extended technical information is out of the scope for the purposes of this study, and thus the interested reader is referred to check the original publication for details [16]. In addition, the source code of the algorithm has been made publicly available as open-source, allowing tracking of all the changes and implementation details. The code (implemented using Matlab) can be downloaded from GitHub [29].

2.3. Experimental procedures

All the recordings from the datasets described in Section 2.1. were rescored by the automatic algorithm. For parameter configuration, two separated and relatively size-reduced datasets, namely HMC-22 and SHHS1-26, were used. Using this approach it is possible to keep the parameterization independent of the testing data (HMC-S, HMC-M, and SHHS2) therefore allowing the possibility to evaluate the generalization capabilities of the algorithm. HMC-22 was used for the validation of an earlier version of our algorithm and it is composed of 22 in-hospital PSG recordings gathered from the HMC database. More detailed description of the dataset and the related validation process can be found in [16]. The SHHS1-26 dataset, on the other hand, is composed of 26 ambulatory PSG recordings gathered from the SHHS-1 study. This dataset was used to validate alternative EEG arousal detection approaches in the past, which are described in detail in [14] and in [15]. Subsets of SHHS1-26 were used as well to validate different machine learning-based approaches described in [12] (n = 20) and in [13] (n = 10).

Reference derivations for automatic EEG arousal analysis vary per dataset due to differences between the respective clinical montages. Specifically, for HMC datasets a C4/M1 EEG with bipolar submental EMG configuration was used on the case of HPSG recordings, while for APSGs the Cz/O2 EEG derivation was used instead. In both cases a single-channel modified lead II ECG derivation was used as reference for the analysis, and for the removal of ECG artifacts from the EEG and the EMG signals [16] [28]. The sampling frequency was 256 Hz for all the signals. Out of the two central EEG derivations available for SHHS recordings [21], the C3/A2 channel was used, together with the default bipolar submental EMG trace. Bipolar-lead ECG was sampled at 125 Hz for SHHS-1, and at 250 Hz for SHHS-2 recordings, and it was used analogously for ECG artifact removal purposes.

The analysis included the automatic rescoring of all EEG arousals, while the remaining (non-EEG arousal) expert annotations were left intact. This includes, for example, the "lights out" and "lights on" markers, and the hypnogram annotations, used respectively to determine the valid scoring and the sleeping periods.

To avoid bias of the analysis due to the unbalance in the recording time between in-hospital and ambulatory recordings, only Time In Bed (TIB) periods were further considered for the validation. Specifically, for HMC datasets this period was extracted straightaway from the "lights out" and "lights on" markers available within each of the EDF+ annotation files, set by the scorers while manually reviewing the recordings. For SHHS, such markers were not explicitly available within the file annotations. In this case TIB was calculated using the variables "*stloutp*" (Lights out time) and "*time_bed*" (TIB in minutes from "lights out" to "lights on") available within the SHHS2 metadata (for details see [30]).

Once all the EEG arousals were automatically scored by the algorithm, the validation was performed using two different and complementary approaches. First, event-to-event scoring validation was carried out using a 30 s epoch basis. For this purpose, every EEG arousal event was assigned to a unique epoch, according to the location of its middle point. Using a 2x2 confusion matrix validation metrics for nominal data, namely sensitivity (recall), specificity, precision, F-1 score, and Cohen´s kappa index were then calculated.

Second, and from a clinical perspective, the respective Arousal Index (ArI) scores were calculated and compared per recording. For validation metrics involving numerical data we used both, the Anderson-Darling and the Lilliefors tests, to check the normal distribution hypothesis. In general, statistical testing was conducted using *Matlab* software, and the reference significance level was set at α = 0.05. Correlation coefficients were calculated among the respective automatic and clinical reference ArI scores. Statistical significance for paired differences at the recording level was calculated using the Wilcoxon signed rank test. In addition, the Intraclass Correlation Coefficient (ICC) was used as the default measure of repeatability to examine scoring differences [31]. Specifically, a two-way absolute single-measures ICC variant of the statistic was considered [32], using the implementation available at [33]. ICC results were calculated both in the original scale and after log-transformation of the respective automatic and clinical reference scores. Furthermore, for non-Gaussian distributions repeatability was also examined using Generalized Linear Mixed-Effects Models (GLMMs) with log-link and multiplicative overdispersion modeling [34]. Parametric bootstrapping and Bayesian methods were used for interval estimation, and randomization methods were used for significance testing. Results are provided both on the original and on the link scales. Specifically, the *rptR* package [34] [35] available for the *R* statistical computing language was used for this purpose.

Assessment of the expected inter-scorer reliability

Analysis of the expected level of inter-scorer variability was performed in order to adequately contextualize the results of the automatic scoring. For such purpose a subset of the available PSGs for each dataset was rescored by an independent expert scorer, and the results were compared against the original clinical scorings. Manual rescoring of all the recordings in each dataset, however, was unattainable in practice due to the high number of recordings, and hence to the high associated costs in human resources. Thus a procedure was established to estimate the actual underlying variability using a reduced subset of recordings. The exact procedure is described as follows:

(i) For each dataset the respective distribution of the kappa indices obtained from the previous automatic vs clinical validation was used as reference.

(ii) From this distribution, the five recordings whose associated indices represent the middle of each inter-quartile range, plus the median, were selected as representatives of the whole population. That is, for each dataset, the recordings with kappa scores on the 12.5, 37.5, 50, 62.5 and 87.5 percentiles were used.

(iii) Each of these recordings was then rescored by a dedicated expert scorer (not present during the first scoring round), blinded to the results of the original analysis. Display montages at the rescoring step were configured to match the same conditions as during the original clinical scoring. Note, that during the rescoring, the hypnogram that resulted from the first scoring was available for contextual interpretation. Its modification on the other hand was not allowed. Note as well that respiratory activity traces were omitted from the display for the purposes of EEG arousal scoring.

Analysis of the rescoring results was carried out by calculating the corresponding derived kappa indices, and by confronting them to both the corresponding clinical and automatic results. Further statistical analyses of the corresponding ArI indices were omitted, as with five measurements per dataset, low statistical power of the derived metrics was expected [36].

## 3. Results

Results of epoch-based event-by-event validation are summarized in Tables 2 and 3. In Table 2, and for each dataset, the total number of epochs and the respective validation metrics are accumulated across all the recordings. In Table 3 statistical descriptors are shown by considering the respective per-recording distributions. In general indices do not follow a normal distribution, and thus data is presented using the median and the associated first and third quartiles.

For the SHHS2 dataset Figure 1 shows the corresponding kappa index distributions in relation to the respective signal quality scores (check [26] for details on SHHS quality assessment procedures). Kruskal-Wallis analyses resulted in $p = 0.004$ and $p < 0.001$ respectively for the EEG and chin EMG distributions. Subsequent multiple comparison tests under the Tukey's honest significant difference criterion showed, however, no group differences for the EEG, while for the EMG, only group 1 was significantly different from groups 4 and 5. Unfortunately, quality assessment data were not available for the HMC database to perform a similar analysis.

Statistical analysis of diagnostic indices

The results from the corresponding ArI distribution analyses are shown in Table 4. Distribution descriptors using the mean and standard derivation, as well as the median and the respective interquartile ranges, are shown per dataset. Individual and difference distributions were analyzed showing non-Gaussian distributions in general ($p < 0.01$ in all the cases). For the difference distributions the corresponding Wilcoxon paired test *p*-value is explicitly shown in the last column. In this respect results reject the null hypothesis H$_0$ "median of differences is zero" at α = 0.05, but for the HMC-M dataset ($p = 0.224$). Further tailed analysis shows that differences for HMC-S and for SHHS2 are not significant anymore when assuming a median difference bias of +0.3 ($p = 0.088$ and $p = 0.104$ respectively).

Results of the repeatability analyses are shown in Table 5. The linear correlation coefficient (*r*) and the ICC indices are shown, both for the original and for the log-transformed variables. GLMMs are adequate in the case of non-Gaussian distributions **[34]**, and the corresponding derived indices were similarly calculated in the original and in the latent scales. In both cases a log-link function was used. Notice that reporting repeatability of the transformed variables is the most interesting choice in most of the cases [34] [37]. For completeness, however, here both estimates are reported. In all the cases statistical significance of the respective tests was confirmed ($p < 0.001$ for all the indices).

**Table 2.** Overall results of the event-by-event epoch-based validation on the testing datasets. Sensitivity (Sens), Specificity (Spec), Precision (Prec), F1-score and Cohen´s kappa index are calculated based on the total number of cases in the respective contingency table. For each dataset the total number of epochs are accumulated across all the recordings; TP= True Positives, FP = False Negatives, TN = True Negatives, FN = False Negatives

| Dataset | #Epochs | TP | FP | TN | FN | Sens | Spec | Prec | F1-score | Kappa |
|---|---|---|---|---|---|---|---|---|---|---|
| HMC-S | 207312 | 12492 | 5170 | 180600 | 9050 | 0.580 | 0.972 | 0.707 | 0.637 | 0.600 |
| HMC-M | 236336 | 13130 | 7340 | 205668 | 10198 | 0.563 | 0.953 | 0.641 | 0.600 | 0.559 |
| SHHS2 | 2201487 | 119702 | 41398 | 1928384 | 112003 | 0.517 | 0.979 | 0.743 | 0.610 | 0.573 |

**Table 3.** Distribution descriptors of the per-recording event-by-event validation metrics. Data is shown as Q2 (Q1, Q3) quartiles; Sens = Sensitivity, Spec = Specificity, Prec = Precision

| Dataset | Sens | Spec | Prec | F1-score | Kappa |
|---|---|---|---|---|---|
| HMC-S | 0.608 (0.476, 0.724) | 0.978 (0.966, 0.986) | 0.715 (0.618, 0.804) | 0.643 (0.539, 0.712) | 0.609 (0.494, 0.683) |
| HMC-M | 0.587 (0.438, 0.731) | 0.973 (0.954, 0.987) | 0.667 (0.484, 0.784) | 0.571 (0.489, 0.658) | 0.529 (0.435, 0.621) |
| SHHS2 | 0.509 (0.394, 0.634) | 0.983 (0.973, 0.989) | 0.757 (0.661, 0.824) | 0.590 (0.503, 0.683) | 0.552 (0.461, 0.651) |

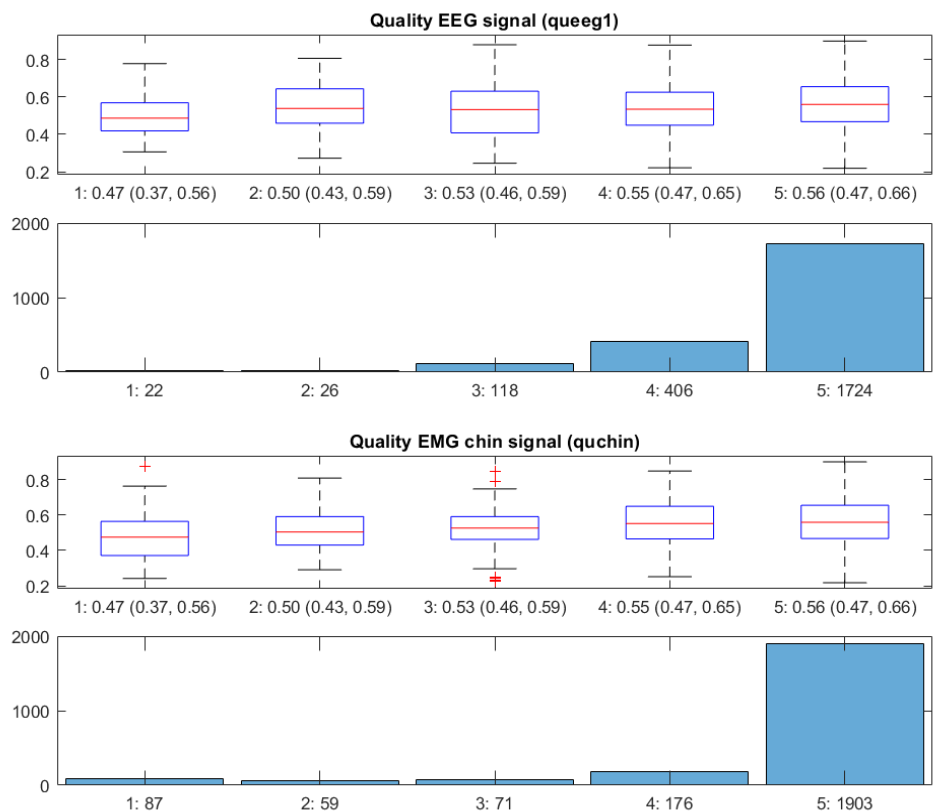

**Figure 1.** Signal quality assessments for the 2296 recordings used for validation in the SHHS2 dataset (automatic vs manual event-by-event validation). Grades were assigned by SHHS scorers according to the SHHS quality assessment procedures. In SHHS2 values vary from 1 (poorest) to 5 (best) and reflect the proportion of sleep time in which the signals were free of artifact; "1": < 25%, "2": 25-49%, "3": 50-74%, "4": 75-94%, "5": > 95%. Upper plot: "Quality of the EEG signal (queeg1)". Lower plot: "Quality of the EMG chin signal (quchin)". On each case the first subplot shows the corresponding kappa distributions per group (numerical values for the median and the inter-quartile ranges are indicated below). The subsequent subplot shows a histogram with the number of recordings involved in the corresponding category.

**Table 4.** Summary of statistical tests for the diagnostic ArI indices (automatic vs clinical reference). *Normality test rejected with p<0.01; **normality test rejected with p<0.001

| | Individual distributions | | | | Difference distribution | | |
|---|---|---|---|---|---|---|---|
| Dataset | Reference | | Auto | | Ref - Auto | | Wilcoxon paired test |
| | Mean±SD | Q2 (Q1, Q3) | Mean±SD | Q2 (Q1, Q3) | Mean±SD | Q2 (Q1, Q3) | p-value |
| HMC-S | 13.32±8.01** | 11.86 (7.68, 16.61) | 12.47±8.06** | 10.74 (7.42, 15.03) | 0.84±5.41** | 0.40 (-1.99, 3.97) | 0.023 |
| HMC-M | 12.45±10.48** | 9.97 (5.46, 15.52) | 12.97±10.14** | 10.56 (6.92, 15.14) | -0.52±6.68* | -0.60 (-4.21, 3.50) | 0.224 |
| SHHS2 | 12.91±7.02** | 11.54 (8.13, 15.97) | 12.56±7.73** | 10.74 (7.37, 15.49) | 0.35±4.89** | 0.44 (-2.07, 3.02) | p<0.001 |

**Table 5.** Repeatability indices calculated over the resulting ArI scores distributions (automatic and clinical reference). CI = Confidence Interval; r = Linear Correlation Coefficient; ICC = Interclass Correlation Coefficient; GLMM = Generalized Linear Mixed-Effects Model. For GLMM n = 100 is used both for parametric bootstrapping and for interval estimation

| Dataset | Metric | Value | 95% CI |
|---|---|---|---|
| HMC-S | r | 0.771 | [0.676, 0.839] |
| | ICC | 0.770 | [0.710, 0.820] |
| | r (log) | 0.693 | [0.624, 0.724] |
| | ICC (log) | 0.694 | [0.617, 0.757] |
| | GLMM (original) | 0.742 | [0.667, 0.806] |
| | GLMM (link-scale) | 0.711 | [0.643, 0.774] |
| HMC-M | r | 0.791 | [0.712, 0.862] |
| | ICC | 0.790 | [0.739, 0.832] |
| | r (log) | 0.646 | [0.552, 0.685] |
| | ICC (log) | 0.648 | [0.568, 0.715] |
| | GLMM (original) | 0.755 | [0.678, 0.799] |
| | GLMM (link-scale) | 0.701 | [0.639, 0.741] |
| SHHS2 | r | 0.780 | [0.762, 0.789] |
| | ICC | 0.780 | [0.764, 0.796] |
| | r (log) | 0.759 | [0.734, 0.769] |
| | ICC (log) | 0.761 | [0.739, 0.780] |
| | GLMM (original) | 0.791 | [0.728, 0.774] |
| | GLMM (link-scale) | 0.761 | [0.697, 0.742] |

Expected inter-rater variability analysis

Results from the expected inter-scorer variability analysis are shown in Table 6. For each recording the representative percentile within its dataset, and the time spent during the manual rescoring, are indicated, respectively, in columns 2 and 3. The resulting kappa indices are also respectively reported for the manual rescoring vs the original clinical annotations (R-C, column 4), the automatic vs the original clinical annotations (A-C, column 5), and the automatic vs the manual rescoring (A-R, column 6) analyses. From these three, R-C is considered to set the reference for the expected levels of (human) inter-scorer variability, as this is the one involving the two independent manual scorings.

By comparing the average kappa values for R-C and A-C similar ranges are noticed (columns 4 and 5: HMC-S 0.594/0.595, HMC-M 0.561/0.523, SHHS2 0.552/0.543). That supports the hypothesis of the automatic algorithm behaving as "one expert more" (i.e. no important differences are evidenced between the human-human and the automatic-human scorings in terms of kappa agreement). There is a slight global increase in the A-R agreements (column 6: 0.602, 0.686, 0.564) as compared to the respective A-C (column 5: HMC-S 0.595, HMC-M 0.523, SHHS2 0.552) and R-C (column 4: HMC-S 0.594, HMC-M 0.561, SHHS2 0.543) agreements. This might be indicative of the automatic algorithm actually behaving more as "the rescoring expert" than as "the original clinical scorers". The effect is more noticeable among the HMC-M recordings. However, it is seldom appreciable for HMC-S and SHHS2 for that to be considered an effective global bias.

Accounting for the differences between the HMC-S and the HMC-M datasets, we might speculate about an expected increment in the variability values on the second case. This is due to the fact that scoring on the HMC-M dataset was performed "purely manual" (i.e. automatic scoring was not used as a pre-scoring step). The intuition behind this hypothesis is simple: a first pass of the automatic algorithm would help to focus the attention of the scorers, contributing both to reduce the time needed for the scoring (revision is limited to checking the results of the automatic analysis), and as a side effect, to increase both the repeatability and the consistency of the scoring criterion (thus, reducing the inter-scorer variability). This hypothesis is only slightly supported by our results with a small relative reduction on the respective R-C agreements (0.594 for HMC-S and 0.561 for HMC-M). Recall, on the other hand, the increased overall agreement achieved by the automatic algorithm in HMC-S as with respect to HMC-M (Table

Table 6. Results of the inter-scorer variability analysis for the HMC-S dataset (A), the HMC-M dataset (B), and the SHHS2 dataset (C). R-C: manual rescoring vs original clinical scorings; A-C: automatic vs original clinical scorings; A-R: automatic scoring vs the manual rescoring

**A**

| Id | Percentile | Time manual rescoring (min) | Kappa index | | |
|---|---|---|---|---|---|
| | | | R-C: Rescoring vs Clinical | A-C: Auto vs Clinical | A-R: Auto vs Rescoring |
| HMCS01 | 12.5 | 40 | 0.449 | 0.437 | 0.324 |
| HMCS02 | 37.5 | 25 | 0.698 | 0.555 | 0.694 |
| HMCS03 | 50 | 40 | 0.628 | 0.609 | 0.687 |
| HMCS04 | 62.5 | 27 | 0.543 | 0.644 | 0.622 |
| HMCS05 | 87.5 | 20 | 0.654 | 0.730 | 0.685 |
| Average | --- | 30.4 | 0.594 | 0.595 | 0.602 |

**B**

| Id | Percentile | Time manual rescoring (min) | Kappa index | | |
|---|---|---|---|---|---|
| | | | R-C: Rescoring vs Clinical | A-C: Auto vs Clinical | A-R: Auto vs Rescoring |
| HMCM01 | 12.5 | 50 | 0.490 | 0.347 | 0.592 |
| HMCM02 | 37.5 | 15 | 0.601 | 0.485 | 0.772 |
| HMCM03 | 50 | 20 | 0.466 | 0.530 | 0.747 |
| HMCM04 | 62.5 | 25 | 0.681 | 0.583 | 0.577 |
| HMCM05 | 87.5 | 20 | 0.567 | 0.672 | 0.743 |
| Average | --- | 26 | 0.561 | 0.523 | 0.686 |

**C**

| Id | Percentile | Time manual rescoring (min) | Kappa index | | |
|---|---|---|---|---|---|
| | | | R-C: Rescoring vs Clinical | A-C: Auto vs Clinical | A-R: Auto vs Rescoring |
| 204977 | 12.5 | 18 | 0.422 | 0.391 | 0.419 |
| 201413 | 37.5 | 32 | 0.605 | 0.509 | 0.607 |
| 202435 | 50 | 30 | 0.574 | 0.552 | 0.603 |
| 203204 | 62.5 | 18 | 0.495 | 0.597 | 0.574 |
| 205545 | 87.5 | 19 | 0.617 | 0.711 | 0.616 |
| Average | --- | 23.4 | 0.543 | 0.552 | 0.564 |

3, median kappa of 0.609 and 0.529 respectively), which might also be explained by the higher expected inter-scorer variability associated to HMC-M. In either case, results are not conclusive on this hypothesis.

Finally, with respect to SHHS2, the expected levels of R-C agreement are the lowest among the three datasets (column 4: HMC-S 0.594, HMC-M 0.561, SHHS2 0.543). We could hypothesize about different contributing factors, such as the fact that the SHHS2 scoring criteria relies on an older version of the standard (ASDA1992 [25]). The use of different montages, or differences due to the different training background of the reference rescoring expert, might have contributed, in addition, to this result. In any case, the slight differences on the respective indices suggest that these factors are not contributing in excess to cause major differences in the expected levels of inter-rater variability. Perhaps, most of the variability can be better explained by the difficulty of the EEG arousal scoring task itself, rather than by such external factors.

Inter-scorer variability reported in the literature

Human inter-scorer variability has been reported for the EEG arousal scoring task in some other works in the literature. Direct comparison between the different studies is however challenging, as the exact methods might differ per study, and exact reproducibility of the experimentation is not always possible.

In a study by Drinnan et al., a comparison of different EEG arousal scorings was carried out from a set of 90 events, and between 14 different European laboratories. A kappa agreement of 0.47 was reported in this study [38]. In a population of 20 patients, with and without obstructive sleep apnea (OSA), Loredo et al. [39] reported an ICC of 0.84 between two human scorers. Significant differences were reported in the same work between the two scorers when comparing the correlation coefficient of the respective ArI differences for two consecutive nights. Pittman et al. [8] calculated the agreement between two human scorers on a dataset of 31 OSA patients, reporting a kappa index of 0.57 using an epoch-by-epoch event validation procedure. An ICC of 0.81 was achieved when comparing the respective ArI scores. More recently, Ruehland et al. [40] reported a median Fleiss kappa of 0.54 (modified for continuous measurements) to estimate the inter-scorer reliability of the EEG arousal scoring task using standard reference montages. They used a dataset of 15 recordings and four different scorers.

In children (n = 36) with and without OSA, Wong et al. [41] calculated the differences between two human scorers resulting in an overall ICC of 0.90 (0.88 for the normal group).

It is worth to mention the study of Whitney et al. [42] as it concerns human variability analysis in the SHHS database. In their study a subset of 30 recordings was used reporting an ICC of 0.54 between three different scorers. The ICC agreement increased to 0.72 when the two most experienced scorers were compared. The intra-scorer variability was also analyzed on 20 out of the 30 recordings, and significant differences were found for two out of the three scorers using a paired t-test among the respective ArI derived indices [42].

The highest inter-rater agreement published in the literature can be found in the study of Smurra et al. [43], who analyzed both the inter- and the intra-scorer reliability for two different scorers on a set of 20 OSA patients. Their analysis was carried out according to two different scoring standards, namely the ASDA1992 and the ULC. In their work Smurra et al. reported an inter-scorer ICC of 0.96 for the ASDA1992, and of 0.98 for the UCL standards. Moreover, no statistical differences were found using one-way ANOVA analysis, when evaluating intra-scorer variabilities between the two scoring references [43].

Results from other automatic approaches in the literature

Some previous works have performed validation procedures based on positive (overlapping) event matchings against the clinician's scorings (Zamora and Tarassenko [5], De Carli et al. [4], and Agarwal [7]). Sensitivity and precision values vary on these studies (0.42-1.00 and 0.57-1.00, respectively) as well as it does the number of PSGs involved (from 2 to 8).

Some other works have carried out epoch-by-epoch validation procedures using time-fixed scoring units. Cho et al. [6] used 1s epochs reporting a sensitivity of 0.75 and a specificity of 0.93 on a set of 6 recordings. Sugi et al. [10], on the other hand, used a 1.28s epoch length for counting positive matches, while a 30s time reference was used for the computation of true negative scores. Reported indices of sensitivity and specificity were 0.82 and 0.88, respectively, on a dataset of 8 recordings. Using a 30s time reference, as in our study, the method of Pacheco and Vaz [3] achieved a sensitivity of 0.88 on selected 2-hour periods from 8 PSG recordings. In another study [11], using a dataset of 20 full recordings and a 30s epoch reference, Shmiel et al. obtained a sensitivity of 0.75 and a precision of 0.77.

Is to be remarked that, at least for the approximations of Zamora and Tarassenko [5], Pacheco and Vaz [3], and Cho et al. [6], the validation was limited to partial pre-selected periods out of the total recording time. In the work of De Carli et al. [4] the standard reference was obtained from the consensus of two human scorers and their own proposed method, which might bias the result. Agarwal [7] mentioned that the configuration of the method was optimized individually for each of the two testing recordings. Therefore, the validation results might be biased as well. The validation procedure described in Sugi et al. [10] presents a similar problem, as 25% of each recording's data were used for training their model. Notice as well that none of the previous works have reported on the respective expected values of inter-scorer reliability for their datasets.

The work of Pittman et al. [8] is more interesting on this regard. Using a dataset of 31 recordings, and following a similar 30s epoch-by-epoch validation procedure, they reported a kappa index of 0.57 between two reference human scorers. Their automatic algorithm achieved kappa values of 0.28 and 0.30, respectively, against each of the two human scorers. In the same work, in addition, a comparison of the ArI derived indices was performed, resulting in a human-human ICC of 0.81, and a human-computer ICC of 0.58 and 0.72.

Comparison with previous automatic approaches from the authors

As introduced in Section 2.3 the authors have attempted automatic EEG arousal scoring in the past following different approaches [16] [15] [14] [13] [12].

Specifically, in [14] and in [15], the full SHHS1-26 dataset was used, and in [16], the HMC-22 dataset was taken as reference for the validation of a preliminary version of the current algorithm. Table 7 shows the results of the current version of the algorithm using the HMC-22 and the SHHS1-26 datasets, together with the results obtained in the original publications. On each case validation procedures were replicated following the exact same conditions as in the original studies, therefore allowing one-by-one direct comparison of the results. The total number of scorable 30s epochs in HMC-22 is of 21826 and in SHHS1-26 of 31080.

To set a baseline in order to evaluate the generalization capabilities of the new version the algorithm with respect the earlier version presented in [16], the original version was re-evaluated using the SHHS1-26 dataset. In addition, we were able to run the algorithm described in [14] (originally validated for the SHHS1-26 dataset only) using the HMC-22 dataset. Similar attempts to re-run the algorithm described in [15] in the HMC-22 dataset were unfortunately unsuccessful, as considerable recoding effort would have been necessary to enable the analysis with the alternative database.

**Table 7.** Performance comparison of previous automatic approaches on the alternative SHHS1-26 and HMC-22 datasets. Following the format used in the original publications, results are shown averaging the respective per-patient indices, calculated over the whole recording time; Sens = Sensitivity, Spec = Specificity, Prec = Precision

| SHHS1-26 | Sens | Spec | Prec | F1-score | Kappa |
|---|---|---|---|---|---|
| Current | 0.581 | 0.979 | 0.739 | 0.634 | 0.597 |
| Approach in [16] | 0.329 | 0.992 | 0.830 | 0.450 | 0.405 |
| Approach in [14] | 0.656 | 0.949 | 0.649 | 0.629 | 0.573 |
| Approach in [15] | 0.810 | 0.878 | 0.560 | 0.660 | 0.580 |
| HMC-22 | | | | | |
| Current | 0.791 | 0.981 | 0.807 | 0.792 | 0.773 |
| Approach in [16] | 0.748 | 0.988 | 0.855 | 0.793 | 0.775 |
| Approach in [14] | 0.531 | 0.962 | 0.555 | 0.506 | 0.470 |

From Table 7 we observe that, the current version of the algorithm considerably outperforms its predecessor [16] in SHHS1-26, while keeping a similar performance in HMC-22. The current version also outperforms, in general, all of the remaining examined approaches, namely [14] and [15]. Given that the SHHS1-26 and the HMC-22 datasets were used during the development phase of the present algorithm, it might be arguable the presence of bias on these results. Nevertheless, the generalization capabilities of the current version of the algorithm have been already proven using the large and independent HMC-S, HMC-M and SHHS2 datasets. Thus, the improvement in the performance is rather interpreted as confirming evidence of the good dataset generalization capabilities of the updated version.

In fact, a lack-of-generalization effect can be observed regarding the results of the algorithm described in [14], showing a decay in the performance when reexamined in HMC-22. Hence, this result supports the hypothesis that the original performance for the method described in [14], evaluated in the SHHS1-26 dataset, included, at least, a component of database-specific overfitting. Overall, the results show that the previously reported algorithms ultimately do not generalize well when confronted with a change in the database source.

## 4. Discussion

This is the largest validation of an automatic EEG arousal scoring algorithm carried out up to date. One of the problems that delays the implementation of automatic scoring systems in the clinical routine, is the difficulty that these algorithms find to preserve their performance outside the controlled experimental environment. Approximations have been reported in the literature showing promising results, but usually validations are restricted to a few, mostly local, and private recordings. Moreover, experimentations are often carried out under controlled or idealized conditions. Closing this gap involves giving proof of the real generalization capabilities when confronting large and heterogeneous databases.

Sources of variability challenging the generalization capabilities of this kind of algorithms are diverse. Among others, different databases involve different signal acquisition and digitalization methods, different population characteristics, and different expert interpretations. Moreover, the latter is not exclusively influenced by differences on the expert's training or background: even when restricting the scoring to the very same recording, human subjectivity still contributes to differences on account of the so-called intra- and inter-rater effects. In consequence, it is fundamental to contextualize the performance results of the algorithm in connection with the (database-specific) expected levels of human scoring variability (or agreement). A fact which, despite some very few exceptions [8], is barely reported among the validation studies in the literature.

The performance of our algorithm was analyzed across large patient samples using both, our own sleep center recordings, and a public external source, namely the SHHS database [17] [21]. PSG recordings out of the HMC database were further organized into different, more specific datasets (HMC-S and HMC-M). A working hypothesis here was to assess possible performance or inter-scorer variability differences, when confronting automatic and clinical results obtained in the context of a semi-automatic approach (using automatic scoring first, then reviewing the results manually) with the results obtained using the classical (manual scoring only) reviewing approach. Even though some trend was depicted in our results for the HMC dataset, evidence was not conclusive on supporting this hypothesis.

Specifically, expected levels of human agreement have been estimated in all the three cases ($\kappa$ = 0.594 HMC-S, $\kappa$ = 0.561 HMC-M, $\kappa$ = 0.543 SHHS2) with our algorithm obtaining comparable levels of performance (see Tables 2, 3 and 6). Therefore, we conclude that our algorithm behaves as one expert more, showing generalization capabilities comparable to the respective expected levels of human agreement. Literature studies which have assessed the agreement between clinical experts on an epoch-by-epoch EEG arousal scoring task have reported kappa indices in the range 0.47–0.57 [38] [8] [40]. Apparently this range is consistent with the values obtained for the datasets used in this study.

When considering ArI agreement in terms of ICC, literature in general is less consistent, with inter-rater agreement varying widely in the range 0.54-0.98 [2] [42] [39] [8] [41] [43]. Although ICC is an adequate statistic to quantify rater (human or automatic) variability, comparison of the different results across the literature is not straightforward. In particular (and leaving aside the earlier mentioned sources of variability) none of the previous publications have clearly specified the exact ICC variant [31] being used for their calculations. The problem is not restrictive of this particular domain [37], and similarly, deviations from normality (although frequent in practice) are usually non-adequately addressed [34]. In the sake of reproducibility, and to increase across-literature comparability, we have tried to overcome these specific limitations in our study. For this purpose we have referenced to the specific procedures, and reported different "flavors" of ICC (among some other repeatability measures) in Table 5.

That being said, we cannot avoid contrasting our automatic-human repeatability scores in SHHS2 (ranging generally in 0.759-0.791) with the values reported by Whitney et al. on a set of 30 recordings for the SHHS database (ICC ranging 0.54-0.72 [42]). Besides the uncertainty regarding the specific ICC version used in Whitney et al., any conclusion derived from such tentative comparison should take into account as well that (i) SHHS scoring procedures have been subject to the supervision of a Reading Center [26] (this procedure is usually absent on a clinical routing), and (ii) that guidelines for event scoring in SHHS were based on an older version of the standards (ASDA1992 vs AASM2017). Luckily enough, at least both references agree in the use of a 3 s arousal scoring rule, and in the use of the EMG for the scoring of arousals in REM. Inter-scorer variability has been reported to decrease abruptly when using arousal definitions shorter than 3 s [39] [41], and also (but less significantly) when no EMG derivation is used [2]. A remarkable result in the SHHS2 dataset is that we have obtained robust behavior almost independently of the quality of the associated signals (see Figure 1). Unfortunately, structured quality assessments enabling similar conclusions were not available in the case of the HMC database.

It is difficult to carry out a reliable comparison with other automatic EEG arousal detection approaches in the literature. Previous validation studies are limited to the use of smaller (2-31 recordings) and non-public datasets. Methodology usually differs, and exact reproducibility of the experimentation is not always possible. Moreover, as previously stated, the general lack of assessment of the expected levels of human agreement makes it difficult to interpret the results reported by these approximations. To our knowledge, only Pittman et al. [8] have co-analyzed the respective levels of expected inter-scorer variability when validating their automatic detector. System validation, in this case, showed performance values under the expected levels of human agreement ($\kappa = 0.57$, ICC = 0.81). Specifically, automatic versus human agreement resulted in kappa indices of 0.28 and 0.30, with ICC values of 0.58 and 0.72, respectively for each of the two human scorers involved in the study.

Direct comparison with previously validated approaches was possible when taking as the reference our previous results using the HMC-22 and SHHS1-26 datasets. At this respect we have shown that our current algorithm does keep, or improve, the reference performance over the different approaches, and for the respective datasets. On the contrary, previous approaches are not able to hold their results when confronted with a database different from which the algorithm was originally designed for. Carefully enough, experimental data suggest superior robustness of our approach as with respect to the current state-of-the-art.

An interesting additional comment concerns whether some time improvement can be expected when using the semi-automatic approach in comparison to classical manual scoring. Bringing together data from Tables 6A-6C as the reference, we have estimated the average manual scoring time to be around 25 min (ranging 18-50 min depending on the recording). While systematic assessment of intra-scorer variability is left as future work, we were able to carry out a second review of some of the recordings, using the same expert, but this time using a semi-automatic approach. This second rescoring was performed blinded to the results of the initial analysis, and with a period of more than 2 months in-between. Overall, an average scoring time improvement of around 20-25% was obtained. This means an average of 5 to 6 minutes saving per recording. Notice that the automatic scoring of one full PSG takes about 30 seconds using a normal laptop computer.

To conclude, we would like to allude to the minimalistic nature of the algorithm with regard to the number of signals involved in the analysis. Specifically, our automatic algorithm operates using one EEG and one chin EMG channels. The choice of the specific EEG and EMG derivations used in this study was driven by the availability of the respective database montages. Central EEG derivations (when possible, referenced to the mastoid, and otherwise to the occipital regions) were chosen in HMC to match as close as possible the respective SHHS montages. Performance effects on HMC by the choice of different EEG channels have not been assessed. While we have opted to use an additional ECG trace for the purpose of ECG artifact removal, unpublished data show that, when evaluated on a large patient sample, ECG filtering does not contribute significantly to the overall performance of our algorithm. However, it does successfully address some specific subsets of recordings highly affected by ECG intrusion.

Future work might also explore the extension of the current method to support multichannel EEG. On an earlier study using a different approach, we have shown that further improvement could be expected by combining independent information from different channels [15].

## Acknowledgments

This work has received financial support from the Xunta de Galicia and the European Union (European Social Fund – ESF)

## References

[1] R. Berry, R. Brooks, C. Gamaldo, S. Harding, C. Marcus and B. Vaughn, The AASM Manual for the Scoring of Sleep and Associated Events: Rules, Terminology and Technical Specifications, Version 2.3, vol. 1, Wetchester, IL: American Academy of Sleep Medicine, 2016.

[2] M. Bonnet, K. Doghramji, T. Roehrs, E. Stepanski, S. Sheldon, A. Walters, M. Wise and A. Chesson, "The scoring of arousal in sleep: Reliability, validity, and alternatives," *Journal of Clinical Sleep Medicine,* vol. 3, no. 2, pp. 133-145, 2007.

[3] O. Pacheco and F. Vaz, "Integrated system for analysis and automatic classification of sleep EEG," in *20th Annual International Conference of the IEEE Engineering in Medicine and Biology Society*, Hong Kong, China, 1998.

[4] F. De Carli, L. Nobili, P. Gelcich and F. Ferrillo, "A method for the automatic detection of arousals during sleep," *Sleep,* vol. 22, no. 5, pp. 561-572, 1999.

[5] M. Zamora and L. Tarassenko, "The study of micro-arousals using neural network analysis of the EEG," in *9th International Conference on Artificial Neural Networks (ICANN'99)*, Edinburgh, UK, 1999.

[6] S. Cho, J. Lee, H. Park and K. Lee, "Detection of arousals in patients with respiratory sleep disorders using a single channel EEG," in *27th Annual Conference of the IEEE Engineering in Medicine and Biology Society*, Shanghai, China, 2005.

[7] R. Agarwal, "Automatic detection of micro-arousals," in *27th Annual Conference of the IEEE Engineering in Medicine and Biology Society*, Shanghai, China, 2005.

[8] S. Pittman, M. MacDonald, R. Fogel, A. Malhotra, K. Todros, B. Levy, A. Geva and D. White, "Assessment of automated scoring of polysomnographic recordings in a population with suspected sleep-disordered breathing," *Sleep,* vol. 27, no. 7, pp. 1394-1403, 2004.

[9] U. Malinowska, P. Durka, K. Blinowska, W. Szelenberger and A. Wakarow, "Micro- and Macrostructure of sleep EEG," *IEEE Engineering in Medicine and Biology Magazine,* vol. 25, no. 4, pp. 26-31, 2006.

[10] T. Sugi, F. Kawana and M. Nakamura, "Automatic EEG arousal detection for sleep apnea syndrome," *Biomedical Signal Processing and Control,* vol. 4, no. 4, pp. 329-337, 2009.

[11] O. Shmiel, T. Shmiel, Y. Daga and M. Teicher, "Data mining techniques for detection of sleep arousals," *Journal of Neuroscience Methods,* vol. 179, pp. 331-337, 2009.

[12] D. Alvarez-Estevez and V. Moret-Bonillo, "Identification of electroencephalographic arousals in multichannel sleep recordings," *IEEE Transactions on Biomedical Engineering,* vol. 58, no. 1, pp. 54-63, 2011.

[13] D. Alvarez-Estevez, N. Sánchez-Maroño, A. Alonso-Betanzos and V. Moret-Bonillo, "Reducing dimensionality in a database of sleep EEG arousals," *Expert Systems with Applications,* vol. 38, no. 6, pp. 7746-7754, 2011.

[14] D. Alvarez-Estevez, "Diagnosis of the sleep apnea-hypopnea syndrome: a comprehensive approach through an intelligent system to support medical decision," 2012.

[15] I. Fernández-Varela, E. Hernández-Pereira, D. Alvarez-Estevez and V. Moret-Bonillo, "Combining machine learning models for the automatic detection of EEG arousals," *Neurocomputing,* vol. 268, pp. 100-108, 2017.

[16] I. Fernández-Varela, D. Alvarez-Estevez, E. Hernández-Pereira and V. Moret-Bonillo, "A simple and robust method for the automatic scoring of EEG arousals in polysomnographic recordings," *Computers in Biology and Medicine,* vol. 87, pp. 77-86, 2017.

[17] S. Quan, B. Howard, C. Iber, J. Kiley, F. Nieto, G. O'Connor, D. Rapoport, S. Redline, J. Robbins, J. Samet and P. Wahl, "The Sleep Heart Health Study: design, rationale, and methods," *Sleep,* vol. 20, no. 12, pp. 1077-1085, 1997.

[18] B. Kemp and J. Olivan, "European data format 'plus' (EDF+), an EDF alike standard format for the exchange of physiological data," *Clinical Neurophysiology,* vol. 114, pp. 1755-1761, 2003.

[19] The National Sleep Research Resource, "The National Sleep Research Resource," [Online]. Available: http://sleepdata.org. [Accessed 2018].

[20] D. Dean, A. Goldberger, R. Mueller, M. Kim, M. Rueschman, D. Mobley, S. Sahoo, C. Jayapandian, L. Cui, M. Morrical, S. Surovec, G. Zhang and S. Redline, "Scaling up scientific discovery in sleep medicine: the National Sleep Research Resource," *Sleep,* vol. 39, no. 5, pp. 1151-1164, 2016.

[21] "Sleep Health Heart Study at NSRR," [Online]. Available: https://sleepdata.org/datasets/shhs/. [Accessed 2018].

[22] S. Redline, M. Sanders, B. Lind, S. Quan, C. Iber, D. Gottlieb, W. Bonekat, D. Rapoport, P. Smith and J. Kiley, "Methods for obtaining and analyzing unattended polysomnography data for a multicenter study. The Heart Health Research Group," *Sleep,* vol. 21, no. 7, pp. 759-767, 1998.

[23] A. Rechtschaffen and A. Kales, A manual of standardized terminology, techniques and scoring system of sleep stages in human subjects, Los Angeles, 1968.

[24] C. Iber, S. Ancoli-Israel, A. Chesson and S. Quan, The AASM manual for the scoring of sleep and associated events: rules, terminology and technical specifications, Westchester, IL, 2007.

[25] The Atlas Task Force of the American Sleep Disorders Association, "EEG arousals: scoring rules and examples," *Sleep,* vol. 15, no. 2, pp. 173-184, 1992.

[26] Case Western Reserve University, "Sleep Heart Health Study: Reading center manual of operations," Case Western Reserve University, Cleveland, Ohio, 2002.

[27] "Sleep Heart Health Study. Montage and sampling rate information SHHS2," [Online]. Available: https://sleepdata.org/datasets/shhs/pages/5-montage-and-sampling-rate-information-shhs2.md. [Accessed 2018].

[28] D. Alvarez-Estevez, I. van Velzen, T. Ottolini-Capellen and B. Kemp, "Derivation and modeling of two new features for the characterization of rapid and slow eye movements in electrooculographic sleep recordings," *Biomedical Signal Processing and Control,* vol. 35, pp. 87-99, 2017.

[29] I. Fernández-Varela and D. Alvarez-Estevez, "GitHub - arousals-detection," 2018. [Online]. Available: https://github.com/bigsasi/arousals-detection.

[30] "SHHS dataset files at NSRR," [Online]. Available: https://sleepdata.org/datasets/shhs/files/datasets. [Accessed 2018].

[31] P. Shrout and J. Fleiss, "Intraclass correlations: uses in assessing rater reliability," *Psychological Bulletin,* vol. 86, no. 2, pp. 420-428, 1979.

[32] K. McGraw and S. Wong, "Forming inferences about some Intraclass Correlation Coefficients," *Phychological Methods,* vol. 1, no. 1, pp. 30-46, 1996.

[33] A. Salarian, "Intraclass Correlation Coefficient (ICC) at MathWorks File Exchange," [Online]. Available: https://nl.mathworks.com/matlabcentral/fileexchange/22099-intraclass-correlation-coefficient--icc-. [Accessed 2018].

[34] S. Nakagawa and H. Schielzeth, "Repeatability for Gaussian and non-Gaussian data: a practical guide for biologists," *Biological Reviews,* vol. 85, pp. 935-956, 2010.

[35] "rptR: Repeatability estimation for Gaussian and non-Gaussian data," [Online]. Available: http://rptr.r-forge.r-project.org/. [Accessed 2018].

[36] H. Kraemer and A. Korner, "Statistical alternatives in assessing reliability, consistency, and individual differences for quantitative measures: application to behavioral measures of neonates," *Psychological Bulletin,* vol. 83, no. 5, pp. 914-921, 1976.

[37] K. Hallgren, "Computing Inter-Rater Reliability for Observational Data: An Overview and Tutorial," *Tutorials in Quantitative Methods for Psychology,* vol. 8, no. 1, pp. 23-34, 2012.

[38] M. Drinnan, A. Murray, C. Griffiths and G. Gibson, "Interobserver variability in recognizing arousal in respiratory sleep disorders," *American Journal of Respiratory and Critical Care Medicine,* vol. 158, no. 2, pp. 358-362, 1998.

[39] J. Loredo, J. Clausen, S. Ancoli-Israel and J. Dimsdale, "Night-to-night arousal variability and interscorer reliability of arousal measurements," *Sleep,* vol. 22, no. 7, pp. 916-920, 1999.

[40] W. Ruehland, T. Churchward, L. Schachter, T. Lakey, N. Tarquino, F. O'Donoghue, M. Barners and P. Rochford, "Polysomnography using abbreviated signal montages: impact on sleep and cortical arousal scoring," *Sleep Medicine,* vol. 16, pp. 173-180, 2015.

[41] T. Wong, P. Galster, T. Lau, J. Lutz and C. Marcus, "Reliability of scoring arousals in normal children and children with obstructive sleep apnea syndrome," *Sleep,* vol. 27, no. 6, pp. 1139-1145, 2004.

[42] C. Whitney, D. Gottlieb, S. Redline, R. Norman, R. Dodge, E. Shahar, S. Surovec and F. Nieto, "Reliability of scoring respiratory disturbance indices and sleep staging," *Sleep,* vol. 21, no. 7, pp. 749-757, 21.

[43] M. Smurra, M. Dury, G. Aubert, D. Rodenstein and G. Liistro, "Sleep fragmentation: comparison of two definitions of short arousals during sleep in OSAS patients," *European Respiratory Journal,* vol. 7, pp. 723-727, 2001.

[44] R. Thomas, "Arousals in sleep-disordered breathing: patterns and implications," *Sleep,* vol. 26, no. 8, pp. 1042-1047, 2004.

[45] D. 't Wallant, V. Muto, G. Gaggioni, M. Jaspar, S. Chellappa, C. Meyer, G. Vandewalle, P. Maquet and C. Phillips, "Automatic artifacts and arousals detection in whole-night sleep EEG recordings," *Journal of Neuroscience Methods,* vol. 258, pp. 124-133, 2016.